\documentclass[12pt]{article}
\begin{document}
\begin{center}
{\huge \bf Renormalizable and unitary Lorentz invariant
model of quantum gravity} \\[10mm]
 S.A. Larin \\ [3mm]
 Institute for Nuclear Research of the
 Russian Academy of Sciences,   \\
 60-th October Anniversary Prospect 7a,
 Moscow 117312, Russia
\end{center}

\vspace{30mm}
\begin{abstract} 
We analyze the $R+R^2$ model of quantum gravity 
where  terms quadratic
in the curvature tensor are added to the General Relativity action. 
This model was recently proved to be a self-consistent quantum theory of gravitation,
being both renormalizable and unitary.  
The model
can be made practically indistinguishable from General Relativity
at astrophysical and cosmological scales by the proper choice of parameters. 
\end{abstract}

\newpage

\section{Introduction}
%\label{intro}

Creation of  quantum  gravity still remains
a prominent task of modern
physics. 

The problem is due to well known perturbative
non-renormalizability of Einstein General Relativity.
In the work \cite{h} it was shown by direct calculations that 
at the one loop level  General Relativity
is renormalizable without matter fields but
becomes unrenormalizable
after inclusion of matter fields.
Then also by explicit calculations it was demonstrated \cite{gsv} that General Relativity
is non-renormalizable at the two-loop level even without matter fields.

{We mean here that the theory is perturbatively non-renormalizable.
There are of course Quantum Field Theory examples of  perturbatively non-renormalizable
theories that lead to clear calculable predictions, such as for example
the non-linear sigma model above two dimensions.
We will work within perturbation theory in the present paper
and will not further consider non-perturbative aspects.

In \cite{st} renormalizability of the 
$R+ R^2$-theory was proved. The proof used a specific covariant
gauge for
simplicity. For general gauges an assumption was made
that ultraviolet divergences have  the so called cohomological
structure. This hypothesis was proved for a
class of background gauges in the work \cite{sib}. Hence we consider  
renormalizability of $R+R^2$ gravity
with four derivatives of the metric as well established. 

We will also call this model briefly as
quadratic gravity. 

But in the works \cite{st,st2} it was also stated that quadratic gravity 
is not physical because it violates  unitarity  
or causality. So this model was commonly considered
as unphysical.

Quite recenyly quadratic quantum gravity was proved to be in
fact unitary \cite{l}. 
Thus the $R+R^2$ model
is a  candidate for the quantum theory of  gravitation.

In the present paper we discuss in detail the exact form of the Lagrangian of quadratic  gravity,
the questions of unitarity, stability of the vacuum state and the behavior of the model
at astrophysical and cosmological scales.

\section{Main part}
%\subsection{Definitions and notations}
We consider the relativistic $R+R^2$ action  
including all  terms quadratic in the
Riemann tensor $R_{\mu\nu\lambda\rho}$ and its simplifications
\begin{equation}
\label{action}
S_{sym}=\int d^{D}x \mu^{-2\epsilon} \sqrt{-g} \left(-M_{Pl}^2 R 
+\alpha R_{\mu\nu}R^{\mu\nu}
+\beta R^2+\delta R_{\mu\nu\rho\sigma}R^{\mu\nu\rho\sigma}
+M_{Pl}^2 \Lambda \right),
\end{equation}
here the $R$-term is the Einstein-Hilbert Lagrangian.
The $\Lambda$-term  is not essential in perturbation theory 
which we consider.

$M_{Pl}^2=1/(16\pi G)$ is the Planck mass squared,
$R_{\mu\nu\rho\sigma}$ is the Riemann tensor, $R_{\mu\nu}$
is  the Ricci tensor and
$R$ is the Ricci scalar. $\alpha$, $\beta$ and $\delta$ are
coupling constants, $D=4-2\epsilon$ is the dimension of the space-time 
within dimensional regularization \cite{dr}. $\epsilon$ is the
regularizarion parameter
and $\mu$ is the  parameter with the dimension of a mass in dimensional regularization.

The Riemann tensor reads
\begin{equation}
R^{\rho}_{\sigma\mu\nu}=\partial_{\mu}\Gamma_{\nu\sigma}^{\rho}-\partial_{\nu}\Gamma_{\mu\sigma}^{\rho}
+\Gamma_{\mu\lambda}^{\rho}\Gamma_{\nu\sigma}^{\lambda}-\Gamma_{\nu\lambda}^{\rho}\Gamma_{\mu\sigma}^{\lambda},
\end{equation}
here are the Christoffell symbols
\begin{equation}
\Gamma_{\mu\nu}^{\alpha}=\frac{1}{2}g^{\alpha\beta}\left(\partial_{\nu}g_{\mu\beta}
+\partial_{\mu}g_{\nu\beta}-\partial_{\beta}g_{\mu\nu}\right).
\end{equation}

Let us underline that dimensional regularization \cite{dr} 
is presently the only known  continuous (not discrete like lattice)
regularization of ultraviolet  divergences appropriate for perturbative calculations and
preserving gauge invariance of gravity.

The term containing the coupling $\delta$ 
in the Lagrangian (\ref{action}) is usually omitted in the literature, see e.g.
\cite{st,sib,str}. This is  because of
the Gauss-Bonnet  identity
\begin{equation}
\int d^4x  \sqrt{-g}\left( 
R_{\mu\nu\rho\sigma}R^{\mu\nu\rho\sigma}-4R_{\mu\nu}R^{\mu\nu}+R^2\right)=0.
\end{equation}
The identity is valid only in four-dimensional space. But the dimension of the space-time
in dimensional regularization is $4-2\epsilon$. Thus it seems that 
the term with the coupling $\delta$ must be preserved 
in the action to have renormalizability.

From the other side it is in principle possible that
one will invent four-dimensional continuous regularizatiom which preserves
gauge invariance of gravitational Lagrangian. Then
the term with the coupling $\delta$ should be omitted.
The number of coupling constants in the Lagrangian
most probably should not depend on the choice of regularization.
In this case the term with the coupling $\delta$ should be omitted
in dimensional regularization also.
The point can be checked with direct calculations of counterterms
of the Lagrangian. To establish the full picture it is most probably
necessary to perform two-loop calculations, as it was with the 
establishing {\bf perturbative} non-renormalizability of pure gravity mentioned in
the introduction.  Corresponding calculations are rather involved
even at the one-loop level. This is a subject
for a seperate publication. 

{It should be also mentioned that there is the known Regge-Wheeler lattice 
regularization which preserves
a form of lattice diffeomorphism invariance.

We will work within perturbation theory. Thus
a linearized theory is considered around the flat space metric
\begin{equation}
g_{\mu\nu}=\eta_{\mu\nu}+h_{\mu\nu},
\end{equation}
here the convention in four dimensions is
$\eta_{\mu\nu}=diag(+1,-1,-1,-1)$. 
Within dimensional regularization  $\eta_{\mu\nu}\eta^{\mu\nu}=D$.
Indexes are raised and lowered by means of
the tensor $\eta_{\mu\nu}$.

Gauge transformations of gravity are generated by diffeomorphisms
$x^{\mu} \rightarrow x^{\mu}+\zeta^{\mu}(x)$ and have the form
\begin{equation}
h_{\mu\nu}\rightarrow h_{\mu\nu}+\partial_{\mu}\zeta_{\nu}+
\partial_{\nu}\zeta_{\mu}
+\left(h_{\lambda\mu}\partial_{\nu}+h_{\lambda\nu}\partial_{\mu}+
(\partial_{\lambda} h_{\mu\nu})\right)\zeta^{\lambda},
\end{equation}
whith arbitrary functions  $\zeta_{\mu}(x)$.

Following standard Faddeev-Popov quantization \cite{fp}, 
see also \cite{sf}, one adds 
to the Lagrangian a gauge fixing term
which can be chosen e.g. in the form
\begin{equation}
S_{gf}=-\frac{1}{2\xi}\int d^D x \mu^{-2\epsilon} F_{\mu}
\partial_{\nu}\partial^{\nu} F^{\mu},
\end{equation}
here $F^{\mu}=\partial_{\nu}h^{\nu\mu}$, $\xi$ is the gauge parameter.
Physical results, of course, do not depend on the allowed  choice
of the form of the gauge fixing term.

One should also add the ghost term 
\begin{equation}
S_{ghost}=\int d^Dx \mu^{-2\epsilon} d^D y \mu^{-2\epsilon} \overline{C}_{\mu}(x)
\frac{\delta F^{\mu}(x)}{\delta \zeta_{\nu}(y)}C_{\nu}(y)=
\end{equation}
\[ 
\int d^D x \mu^{-2\epsilon} \partial^{\nu}\overline{C^{\mu}}\left[\partial_{\nu}C_{\mu}
+\partial_{\mu}C_{\nu}
+h_{\lambda\mu}\partial_{\nu}C^{\lambda}
+h_{\lambda\nu}\partial_{\mu}C^{\lambda}+
(\partial_{\lambda}h_{\mu\nu})C^{\lambda}\right], 
\]
where $\overline{C}$ and $C$ are ghost fields.
Then one gets the generating functional of graviton Green functions 
\begin{equation}
Z(J)=N^{-1} \int d h_{\mu\nu} dC_{\lambda} d\overline{C_{\rho}}
 \exp{\left[i\left(S_{sym} +S_{gf}+S_{ghost}+ \int
d^Dx \mu^{-2\epsilon} J_{\mu\nu} h^{\mu\nu}\right)\right]},
\end{equation}
here $N$ is the normalization factor 
of the functional integral in the usual notation, 
$J_{\mu\nu}$
is as usual the source of gravitons.

We work within perturbation theory, hence one makes the shift of the fields
\begin{equation}
 h_{\mu\nu} \rightarrow M_{Pl} \mu^{-\epsilon} h_{\mu\nu}.
\end{equation}
Perturbative expansion is in  inverse powers of the Plank mass 
or in other words in  powers of the  Newton coupling constant $G \propto 1/M_{Pl}^2$.

Let us obtain the graviton propagator. 
One takes the part of the Lagrangian quadratic in $ h_{\mu\nu}$
and makes the Fourier transform
\[
Q=
\frac{1}{4}\int d^D k~ h^{\mu\nu}(-k)\left[\left(k^2+M_{Pl}^{-2}k^4(\alpha 
+4\delta)\right)P^{(2)}_{\mu\nu\rho\sigma} \right.
\]
\begin{equation}
\label{forma}
+k^2\left(-2+4M_{Pl}^{-2}k^2(\alpha+3\beta+\delta)\right)
P^{(0-s)}_{\mu\nu\rho\sigma}
\end{equation}
\[ \left. 
+\frac{1}{\xi}M_{Pl}^{-2}k^4\left(P^{(1)}_{\mu\nu\rho\sigma}
+2P^{(0-w)}_{\mu\nu\rho\sigma}\right) \right] h^{\rho\sigma}(k),
\]
$P^{(i)}_{\mu\nu\rho\sigma}$ being projectors 
to the spin-2, spin-1 and spin-0 
components of the field $h_{\mu\nu}$: 
\begin{equation}
P^{(2)}_{\mu\nu\rho\sigma}=\frac{1}{2}\left(\Theta_{\mu\rho}\Theta_{\nu\sigma}
+\Theta_{\mu\sigma}\Theta_{\nu\rho}\right)
-\frac{1}{3}\Theta_{\mu\nu}\Theta_{\rho\sigma},
\end{equation}
\begin{equation}
P^{(1)}_{\mu\nu\rho\sigma}=\frac{1}{2}\left(\Theta_{\mu\rho}\omega_{\nu\sigma}
+\Theta_{\mu\sigma}\omega_{\nu\rho}
+\Theta_{\nu\rho}\omega_{\mu\sigma}+\Theta_{\nu\sigma}\omega_{\mu\rho}\right),
\end{equation}
\begin{equation}
P^{(0-s)}_{\mu\nu\rho\sigma}=\frac{1}{3}\Theta_{\mu\nu}\Theta_{\rho\sigma},
\end{equation}
\begin{equation}
P^{(0-w)}_{\mu\nu\rho\sigma}=\omega_{\mu\nu}\omega_{\rho\sigma}.
\end{equation}
Here $\Theta_{\mu\nu}=\eta_{\mu\nu}-k_{\mu}k_{\nu}/k^2$ and 
$\omega_{\mu\nu}=k_{\mu}k_{\nu}/k^2$
are transverse and longitudinal projectors correspondingly.

%The sum of the projectors is the unity
%\begin{equation}
%P^{(2)}_{\mu\nu\rho\sigma}+P^{(1)}_{\mu\nu\rho\sigma}
%+P^{(0-s)}_{\mu\nu\rho\sigma}+P^{(0-w)}_{\mu\nu\rho\sigma}=
%\frac{1}{2}(\eta_{\mu\rho}\eta_{\nu\sigma}
%+\eta_{\mu\sigma}\eta_{\nu\rho}).
%\end{equation}

We note that the expression (\ref{forma}) differs from the similar
expression presented in  \cite{str} by the absence of  $\epsilon$-dependent terms.

To get the graviton propagator $D_{\mu\nu\rho\sigma}$
one inverts the matrix in square brackets of the expression (\ref{forma}):
\begin{equation}
[Q]_{\mu\nu\kappa\lambda}D^{\kappa\lambda\rho\sigma}
=\frac{1}{2}(\delta_{\mu}^{\rho}\delta_{\nu}^{\sigma}
+\delta_{\mu}^{\sigma}\delta_{\nu}^{\rho}).
\end{equation}
Then the propagator has the form
\begin{equation}
D_{\mu\nu\rho\sigma}=\frac{1}{i(2\pi)^D}
\left[
\frac{4}{k^2}\left(\frac{1}{1+M_{Pl}^{-2}k^2(\alpha +4\delta)}\right)
P^{(2)}_{\mu\nu\rho\sigma}
\right.
\end{equation}
\[
-\frac{2}{k^2}\left(\frac{1+2\epsilon\frac{1-M_{Pl}^{-2}k^2(\alpha+4\beta 
 )}{1+M_{Pl}^{-2}k^2(\alpha+4\delta)}}
 {1-\epsilon-M_{Pl}^{-2}k^2\left((2\alpha+6\beta+2\delta)
-\epsilon (\alpha+4\beta)\right)}\right)
P^{(0-s)}_{\mu\nu\rho\sigma}
\]
\[ \left.
+4\xi \frac{1}{ M_{Pl}^{-2}k^4} \left(P^{(1)}_{\mu\nu\rho\sigma}+
\frac{1}{2}P^{(0-w)}_{\mu\nu\rho\sigma}\right)\right].
\]
Then one  performs partial fractioning. The  propagator
takes the forn
\begin{equation}
D_{\mu\nu\rho\sigma}=\frac{1}{i(2\pi)^D}
\left[
4P^{(2)}_{\mu\nu\rho\sigma}\left(\frac{1}{k^2}
-\frac{1}{k^2-M_{Pl}^2/(-\alpha-4\delta)}\right)
\right. 
\end{equation}
\[ 
-2\frac{P^{(0-s)}_{\mu\nu\rho\sigma}}{1-\epsilon}\left(
1+2\epsilon\frac{1-M_{Pl}^{-2}k^2(\alpha+4\beta)}
{1+M_{Pl}^{-2}k^2(\alpha+4\delta)}\right)
\]
\[
\left(\frac{1}{k^2}-\frac{1}{k^2-M_{Pl}^2(1-\epsilon)/
(2\alpha +6\beta+2\delta-\epsilon (\alpha+4\beta))}\right)
\]
\[ \left.
+\frac{4\xi}{M_{Pl}^{-2}k^4}\left(P^{(1)}_{\mu\nu\rho\sigma}+
\frac{1}{2} P^{(0-w)}_{\mu\nu\rho\sigma}\right)\right].
\]

%We want to note that one  pole in the term
%contaning $P^{(0-s)}_{\mu\nu\rho\sigma}$ depends on the 
%dimensional regularization parameter  $\epsilon$. 
%The  residues of both poles in the same term also  depend on
%the parameter $\epsilon$. Thus it is 
%clear that the poles and the residues of the tree
%graviton  propagator do not possess a physical meaning.

In four dimensions one obtains the following  graviton propagator
\begin{equation}
\label{four}
D_{\mu\nu\rho\sigma}=\frac{4}{i(2\pi)^D}
\left[
\frac{P^{(2)}_{\mu\nu\rho\sigma}
-\frac{1}{2}P^{(0-s)}_{\mu\nu\rho\sigma}}{k^2}
-\frac{P^{(2)}_{\mu\nu\rho\sigma}}{k^2-M_{Pl}^2/(-\alpha-4\delta)}
\right.
\end{equation}
\[ \left.
+\left(\frac{1}{2}\right)\frac{P^{(0-s)}_{\mu\nu\rho\sigma}}{k^2-M_{Pl}^2/
(2\alpha +6\beta+2\delta)}
+\frac{\xi}{M_{Pl}^{-2}k^4}\left(P^{(1)}_{\mu\nu\rho\sigma}+
\frac{1}{2} P^{(0-w)}_{\mu\nu\rho\sigma}\right)\right],
\]

We will now consider classical quadratic gravity.
In this case
for a point particle 
having the energy-momentum tensor
$T_{\mu\nu}=\delta^0_{\mu}\delta^0_{\nu}M\delta^3(x)$ 
one gets the gravitational field \cite{st2} 
\begin{equation}
\label{potential}
V(r)=\frac{M}{2 \pi M_{Pl}^2}\left(-\frac{1}{4 r}+\frac{e^{-m_{2} r}}{3 r}
-\frac{e^{-m_0 r}}{12 r}\right).
\end{equation}
$m_2^2= M_{Pl}^2/(-\alpha-4\delta)$
and  $m_0^2=M_{Pl}^2/(2\alpha+6\beta+2\delta)$ are squared masses
correspondingly of  massive spin-2 and spin-0 gravitons.
Cupling constants $\alpha, \beta$ and $\delta$ can be chosen to obtain
positive masses. 
In
\cite{st,st2} it was noted that  masses can be chosen large enough 
to have an agreement with experiments.

Our propagator (\ref{four}) reproduces 
the expression (\ref{potential}). One can see it by means of the calculation of
the tree level Feynman diagram corresponding to an exchange of two
point-like particles by a graviton. 

The graviton propagator in the work \cite{st} 
dos not produce the expression (\ref{potential}).
It contains some  technical errors.  
To see this one puts in the $R+R^2$ Lagrangian  coupling constants equal to zero
except the Newton coupling. The Lagrangian is reduced then
to General Relativity. Hence the graviton
propagator should also be reduced to one of General Relativity:
\begin{equation}
\label{propgr}
D_{\mu\nu\rho\sigma}(k)=\frac{1}{i(2\pi)^4}\frac{1/2\eta_{\mu\rho}\eta_{\nu\sigma}
+1/2\eta_{\mu\sigma}\eta_{\nu\rho}-1/2\eta_{\mu\nu}\eta_{\rho\sigma}+terms \propto k}{k^2},
\end{equation}
where the gauge condition with $\xi=0$ is taken for simplicity.

Our propagator (\ref{four}) reproduces the propagator (\ref{propgr})
in this limit. The propagator of the work \cite{st}
has
the factor 1 instead of 1/2 in the third term of the numerator of (\ref{propgr}) 
in the corresponding limit.

The second term in the expression (\ref{four}) for the graviton propagator has the 
non-standard minus sign.
Hence one considers it as the massive spin-2 ghost.
For renormalizability of quadratic gravity one must shift
poles of all propagators in Feynman diagrams in the same way
$k^2 \rightarrow k^2 +i0$. Hence the spin-2 ghost must be considered
as a state with negative metric \cite{st}.
That is why  violation of
either unitarity or causality within the $R+R^2$ model
was claimed in \cite{st,st2}.

But this massive spin-2 ghost is unstable. It  unavoidably decays in two or more
physical massless gravitons. The width of the decay is
small. However independently of the value
of this decay width  spin-2 ghost particles do not appear 
as asymptotic states
of the $S$-matrix elements. Only physical gravitons appear
as external particles of the $S$-matrix amplitudes. Thus one concludes
that unitarity is preserved
in the $R+R^2$ model.

There is a statement about instability of theories with ghosts , i.e. their
Hamiltonians are unbounded from below and they
do not have stable vacuum states. This question was raised in \cite{pu} within
Quantum Mechanics, see also
\cite{sm} for the brief review.
But this statement is proved only for Quantum Mechanical systems.
Quantum Field Theory is quite a different story and renomalizability 
plus unitarity is enough to have a consistent theory.

To see this let us consider the graviton propagator in the 
opeator formalism:
\begin{equation}
D_{\mu\nu\rho\sigma}(x-y)=\frac{\delta^2}{\delta J_{\mu\nu}(x)\delta J_{\rho\sigma}(y)}
Z(J)=<0|T\left[h_{\mu\nu}(x)h_{\rho\sigma}(y)\right]|0>.
\end{equation}
One transforms it to the momentum space and inserts the sum
over the complete set of momentum eigenstates between two 
graviton fields. The states with negative norms in the sum
have the extra factor $-1$. It gives the negative residue
for the massive spin-2 ghost pole.

There is another way to produce the negative residue for 
the spin-2 ghost. One can prescribe negative energy to this ghost.
The expansion of the graviton fields into the creation and annihilation
operators produces normalization factors $1/\sqrt{-2k_0}$.
This is the reason for the negative residue for the spin-2 ghost.
In this case of negative energy the Hamiltonian would be indeed
unbounded and the vacuum state would be unstable.

But as it was mentioned above, one should choose the variant
with negarive metric in order to have renormalizability in the theory \cite{st}.
Thus one has the consistent theory with the stable ground state.
There are no reasons for a Hamiltonian to be unbounded from below
if there are no states with negative energies.

It should be mentioned that the $S$-matrix by construction automatically
 satisfies the unitarity relation 
\begin{equation}
\label{uni}
 S^{+}S=1
\end{equation}
in theories having Hermitian Lagrangians \cite{bs}.

To see it one considers the $S$-matrix 
in the operator formalism
\begin{equation}
S=T\left(e^{ i \int L(x) dx} \right).
\end{equation}
One introduces a function $g(x)$ having the values in the interval
$(0,1)$. This function describes intensity of interactions. 
Interactions 
are switched off if $g(x)=0$. If $g(x)=1$ then interactions are switched on.
Interactions are switched on partly if $0<g(x)<1$.
One substitutes the product $L(x)g(x)$ for the Lagrangian $L(x)$.
%one gets interactions switched on with intensity $g(x)$.
The $S$-matrix becomes the functional
\begin{equation}
S(g)=T\left(exp~~ i \int L(x)g(x) dx \right).
\end{equation}
One splits the interaction region characterised by the function
$g(x)$ into an infinitely large number
of infinitely thin segments $\Delta_i$ using the space-like surfaces $t=const$. 

Then one gets
\begin{equation}
S(g)=T\left(exp~~ i \int L(x)g(x) dx \right)=
T\left(exp~~ i\sum_j \int_{\Delta_j} L(x)g(x) dx \right)=
\end{equation}
\[
T\left(\prod_j exp~~ i\int_{\Delta_j} L(x)g(x) dx \right).
\]
$S(g)$ is defined as the limit
\begin{equation}
\label{product}
S(g)=\lim_{\Delta_j \rightarrow 0}T\left(\prod_j\left(1+ 
i\int_{\Delta_j} L(x)g(x) dx \right)\right).
\end{equation}
The r.h.s. of (\ref{product}) is a  product
taken in the chronological order of the segments $\Delta_j$.
Each factor in this product 
is unitary up to small terms of higher orders for sufficiently small $\Delta_j$.
These higher orders can be neglected in the considered limit.
Hence the whole product
is unitary. Unitarity of $S(g)$ and of the matrix
\begin{equation} 
S=\lim_{g(x)\rightarrow 1} S(g) 
\end{equation}
is proved.

Sometimes one understands the following thing  under unitarity.
One derives from  (\ref{uni}) the famous optical theorem stating that imagenary part
of an amplitude of some forward scattering coincides up to a factor
with the corresponding total annihilation crosssection
\begin{equation}
\label{op}
Im <i|T|i>= \frac{1}{2}\sum_n<i|T^{+}|n><n|T|i>,
\end{equation}
where $|i>$ is the scattering state, $T$ is the scattering matrix: $S=1+iT$,
and one assumes that all physical states $|n>$ form a complete set in the theory
\begin{equation}
\label{complete}
\sum_n|n><n|=1.
\end{equation} 
From the other side one can calculate  $Im <i|T|i>$ directly from Feynman diagrams
using Cutcosky cuts. Then one assumes that the result should coincide with (\ref{op}).
But if it does not happen it does not mean violation of unitarity.
It only means that physical states in the theory do not form a complete
set (\ref{complete}) and the complete set is formed by physical plus unphysical states.

Unitarity of theories with
negative metric states was previously  considered 
in \cite{lee,cut}, see also references therein. 
Question of causality were also considered there.

We would like to note that the tree level graviton 
propagator (\ref{four}) is
modified by the summation of the chain of one-loop insertions.
As it was already mentioned above the second term of the propagator (\ref{four})
has the  minus sign. Therefore the summation of
the one-loop insertions with the massless graviton
in the loop will shift the pole of the spin-2 ghost from the value
 $k^2= M_{Pl}^2/(-\alpha-4\delta)$ to the complex value
$k^2= M_{Pl}^2/(-\alpha-4\delta)-i\Gamma$. Here $\Gamma$
is the width of the spin-2 ghost decay  into the pair
of massless physical gravitons.
This complex pole is located on the
unphysical Riemann sheet. It is analogous to the known
virtual level of the neutron-proton system  
with opposite spins of nucleons \cite{ll}.

It should be noted that one loop corrections in quadratic gravity
were studied in \cite{b}.

We would like to underline  that we consider not pure $R^2$ theory but
the $R+R^2$ theory where the $R^2$ terms are added to the Einstein-Hilbert
Lagrangian. Gravitational constants $\alpha$, $\beta$ and $\delta$ 
of these terms in the Lagrangian can be chosen suffisiently small to ensure
that quadratic gravity
will be practically indistinguishable from General Relativity
at astrophysical and cosmological scales.
This is independently of the discussed above in detail question
wether the coupling $\delta$ is exactly zero or not.
The $R^2$ terms are introduced only to have renormalizability
of quadratic gravity which is valid in particular for arbitrary small 
couplings $\alpha$, $\beta$ and $\delta$.

We have analyzed  only purely gravitational $R+R^2$ action. 
The  inclusion of the matter fields in the Lagrangian is
straightforward and does not change conclusions.

\section{Conclusions}
We have proved  unitarity of  quantum gravity with 
the $R+R^2$ action. This model was previously shown to be renormalizable 
in the work \cite{st}. The parameters of quadratic gravity can be adjusted
to ensure
that the theory
will be practically indistinguishable from General Relativity
at astrophysical and cosmological scales.

One can conclude that the $R+R^2$ model is an appropriate
candidate for the fundamental quantum theory of gravity.

\section{Acknowledgments}
The author is grateful 
to the collaborators of the Theory Division of the Institute for Nuclear Research
for valuable discussions.

\end{document}